\documentclass[
    ,final            
  ]
  {aipproc}

\layoutstyle{6x9}

\begin{document}

\title{Production of one and two $c \bar c$ pairs at LHC}


\author{Antoni Szczurek}{
  address={Institute of Nuclear Physics PAN, PL-31-342 Cracow, Poland,\\
           University of Rzesz\'ow, PL-35-959 Rzesz\'ow, Poland}}

\author{Rafa{\l} Maciu{\l}a}{
  address={Institute of Nuclear Physics PAN, PL-31-342 Cracow, Poland}
}


\begin{abstract}
We report on charm production at LHC. The production of single 
$c \bar c$ pairs is calculated in the $k_t$-factorization approach
with different unintegrated gluon distributions.
Examples of transverse momentum distributions for charmed mesons
are presented and compared to recent experimental results from LHC.
Some missing strength is observed for most of UGDFs.
Furthermore we discuss production of two $c \bar c$ pairs within
double-parton scattering (DPS) and single-parton scattering (SPS) mechanisms. 
Surprisingly large cross sections, comparable to single $c \bar c$ pair
production are predicted. We discuss first experimental results
from LHCb collaboration on production of pairs of $D$ mesons of 
the same flavour. 
\end{abstract}

\maketitle


\section{Introduction}
\label{intro}

The cross section for open charm production at the LHC is very large.
Different mesons have been measured recently
\cite{ALICE_charm,LHCb_charm}. Some other 
experiments are preparing their experimental cross sections. 
Different theoretical approaches for heavy quark production are
used in the literature. In the present communication we present
briefly some results for charmed meson production within
$k_t$-factorization approach. A more detailed analysis will be presented
elsewhere \cite{MS2012_Dmesons}. 
Previously we used the $k_t$-factorization approach for charm
production at the Tevatron \cite{LS2006} and for nonphotonic electron 
production at RHIC \cite{LMS2009,MSS2011}.
The $k_t$-factorization approach was also successfully used for beauty
\cite{JKLZ2011} and top \cite{LZ2011} quark (antiquark) inclusive production.

Recently we have made first estimates for the production of two $c \bar
c$ pairs \cite{LMS2011,SS2012}. We have considered both double-parton 
scattering (DPS) mechanism \cite{LMS2011} as well as single-parton scattering 
(SPS) mechanism \cite{SS2012}. Comparison of contributions of both 
mechanisms leads to the conclusion that the production of two 
$c \bar c$ pairs is a favourite place to study and identify 
double-parton scattering effects. Recently the LHCb collaboration
has measured several pairs of $D$ mesons \cite{LHCb_DPS}. We argue that their
measurement confirms large double-parton scattering effects.

\section{Sketch of formalism}
\label{formalism}

In the leading-order (LO) approximation within the $k_t$-factorization approach
the quadruply differential cross section in the rapidity 
of $Q$ ($y_1$), in the rapidity of $\bar Q$ ($y_2$) and in the transverse 
momentum of $Q$ ($p_{1,t}$) and $\bar Q$ ($p_{2,t}$) can be written as
\begin{eqnarray}
\frac{d \sigma}{d y_1 d y_2 d^2p_{1,t} d^2p_{2,t}} =
\sum_{i,j} \; \int \frac{d^2 \kappa_{1,t}}{\pi} \frac{d^2 \kappa_{2,t}}{\pi}
\frac{1}{16 \pi^2 (x_1 x_2 s)^2} \; \overline{ | {\cal M}_{ij \to Q \bar Q} |^2}\\
\nonumber 
\delta^{2} \left( \vec{\kappa}_{1,t} + \vec{\kappa}_{2,t} 
                 - \vec{p}_{1,t} - \vec{p}_{2,t} \right) \;
{\cal F}_i(x_1,\kappa_{1,t}^2) \; {\cal F}_j(x_2,\kappa_{2,t}^2) \; , 
\nonumber  
\end{eqnarray}
where ${\cal F}_i(x_1,\kappa_{1,t}^2)$ and ${\cal F}_j(x_2,\kappa_{2,t}^2)$
are so-called unintegrated gluon (parton) distributions. 

The hadronization is done in the way explained in Ref.\cite{LMS2009}.

\begin{figure}[!h]
\includegraphics[width=5cm]{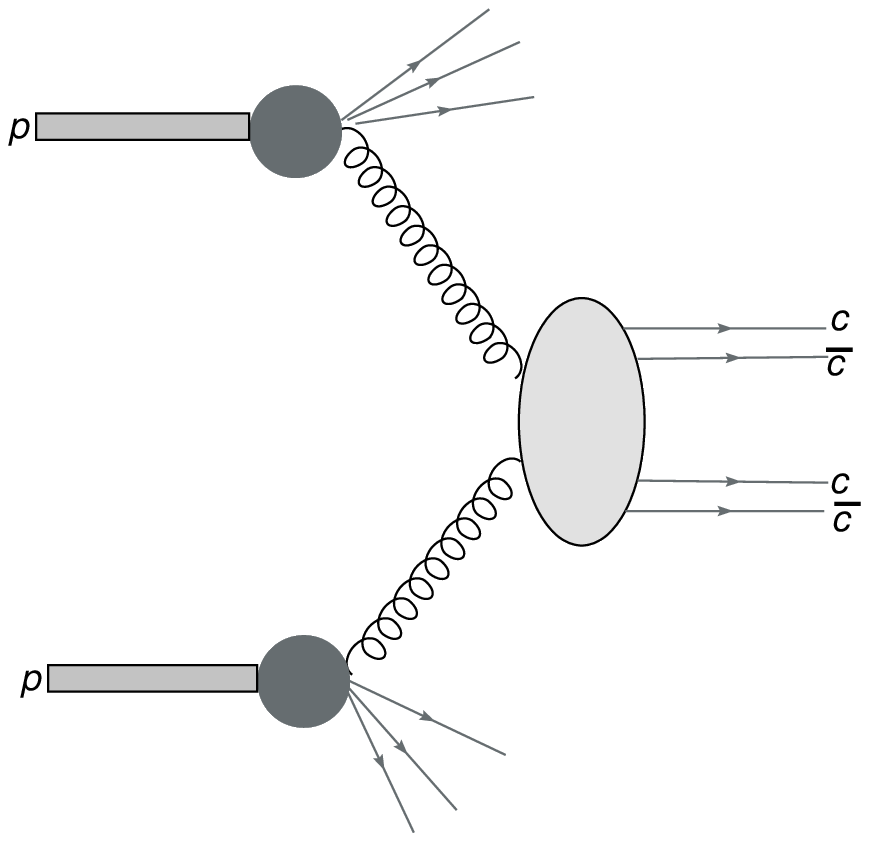}
\includegraphics[width=5cm]{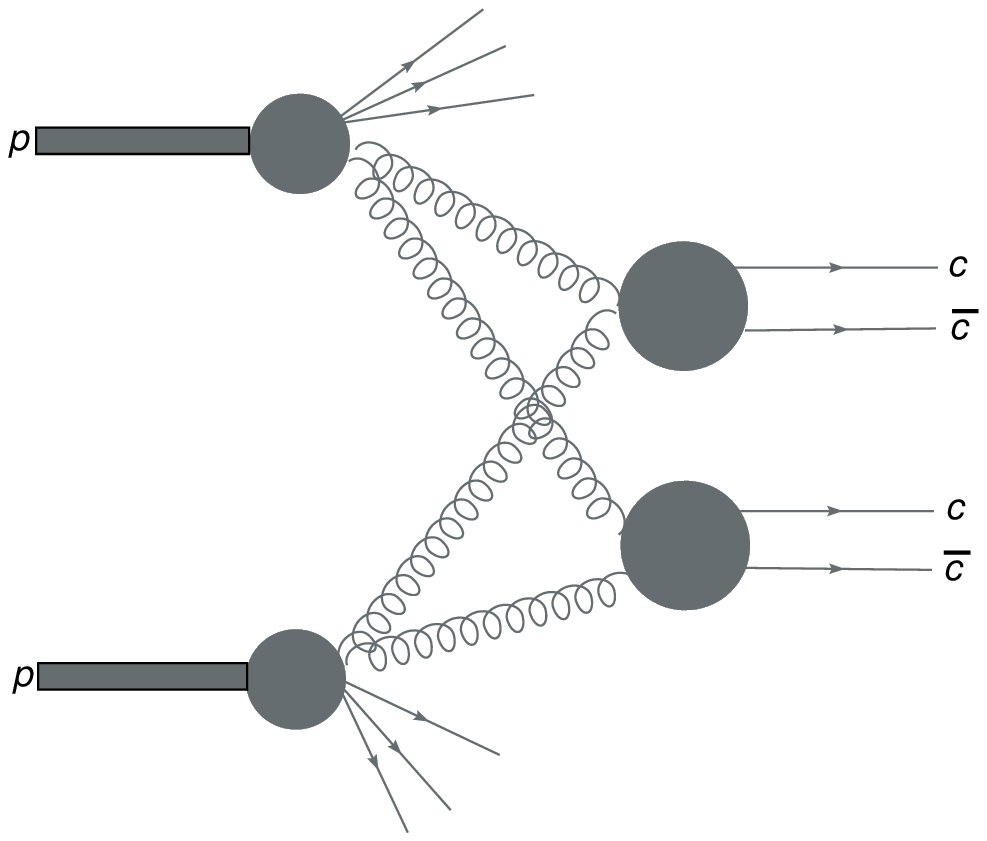}
   \caption{
\small SPS (left) and DPS (right) mechanisms of $(c \bar c) (c \bar c)$ 
production.  
}
\label{fig:mechanisms_ccbarccbar}
\end{figure}

The cross section for differential distribution in a simple
double-parton scattering in leading-order collinear approximation 
can be written as
\begin{equation}
\frac{d \sigma}{d y_1 d y_2 d^2 p_{1t} d y_3 d y_4 d^2 p_{2t}}  \\ =
\frac{1}{ 2 \sigma_{eff} }
\frac{ d \sigma } {d y_1 d y_2 d^2 p_{1t}} \cdot
\frac{ d \sigma } {d y_3 d y_4 d^2 p_{2t}} 
\label{differential_distribution}
\end{equation}
which by construction reproduces the formula for integrated cross
section \cite{LMS2011}.
This cross section is formally differential in 8 dimensions but can be 
easily reduced to 7 dimensions noting that physics of unpolarized
scattering cannot depend on azimuthal angle of the pair or on azimuthal
angle of one of the produced $c$ ($\bar c$) quark (antiquark).
This can be easily generalized by including QCD evolution effects
for double parton distributions \cite{LMS2011}.

Recently we have generalized this approach to the $k_t$-factorization approach 
\cite{MS2012_DPS} where transverse momenta of particles 1 and 2 as well as 
transverse momenta of particles 3 and 4 are not balanced.
This approach generate effectively higher-order corrections.

\section{Results}
\label{results}

\begin{figure}[!h]
\includegraphics[width=6cm]{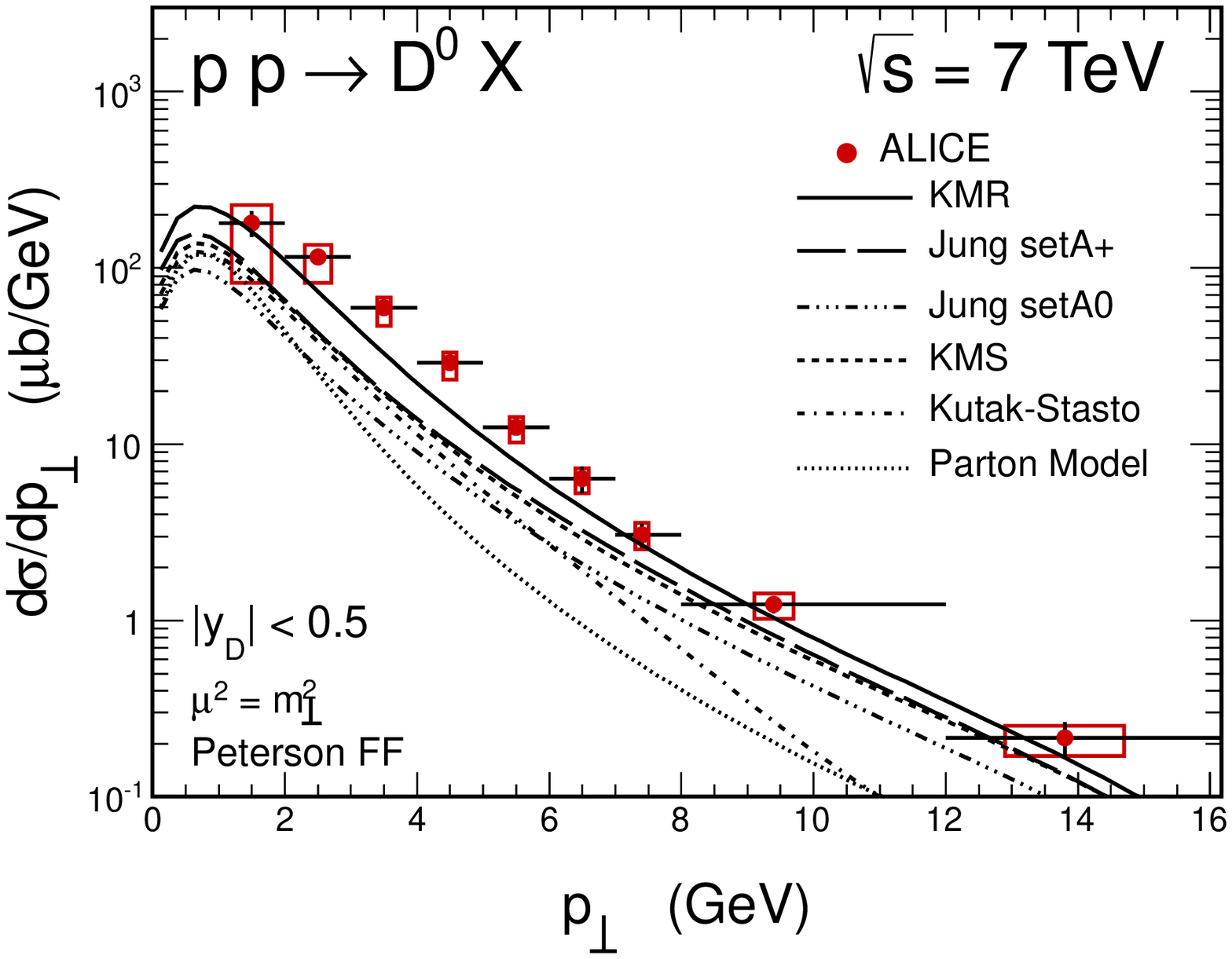}
\includegraphics[width=6cm]{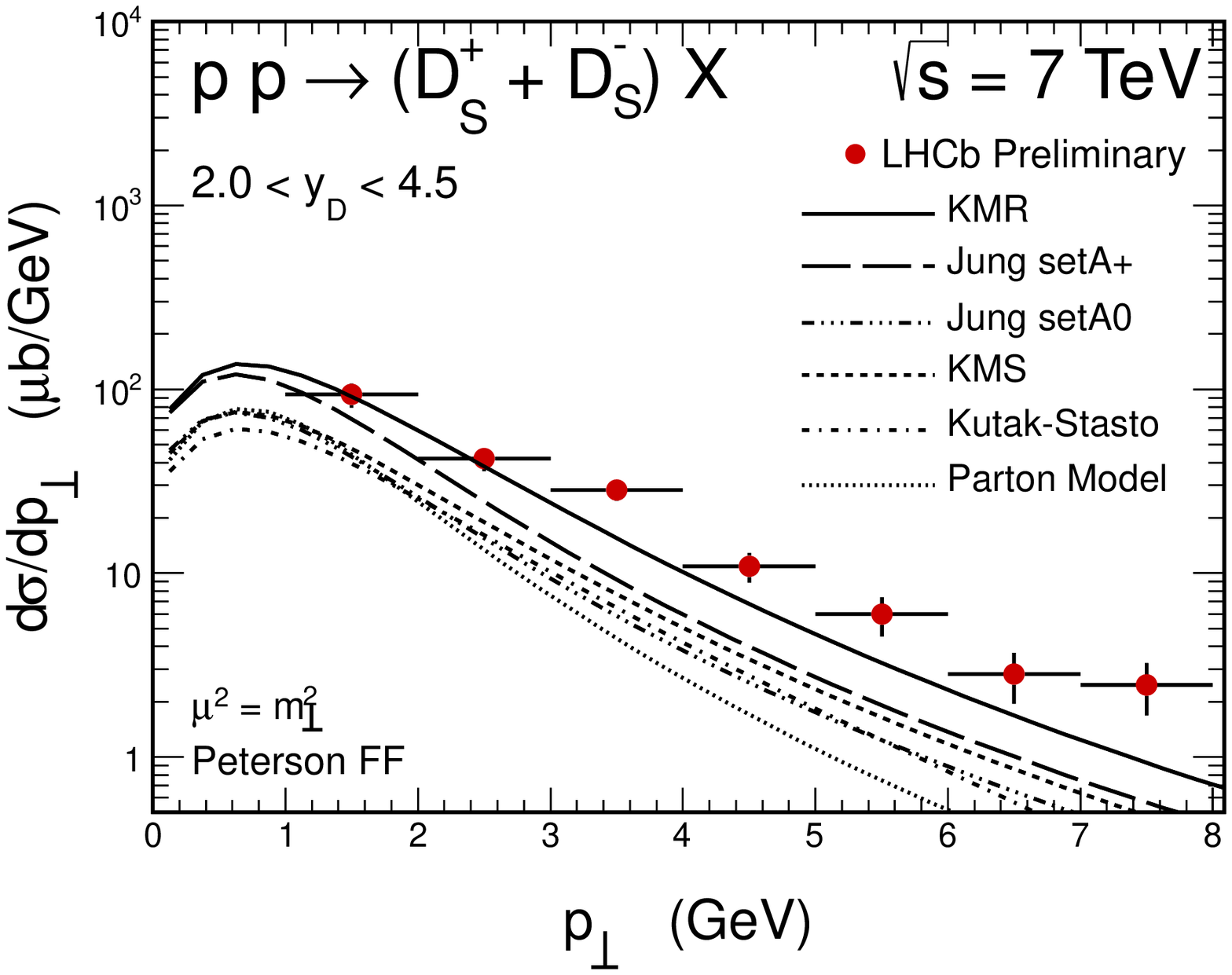}
   \caption{
\small Two examples of transverse momentum distribution of charmed
mesons compared to ALICE (left panel) and LHCb (right panel)
experimental data. The calculations were done for different unintegrated
gluon distributions.
}
 \label{fig:dsig_dpt_mesons}
\end{figure}

In Fig.\ref{fig:dsig_dpt_mesons} we show two examples of transverse
momentum distribution of $D$ mesons. Our results are compared with the
recent experimental data \cite{ALICE_charm,LHCb_charm}. 
Some strength seems to be missing. A possible explanation is 
discussed below.
More distributions will be shown in our future publication 
\cite{MS2012_Dmesons}.

\begin{figure}[!h]
\includegraphics[width=6.0cm]{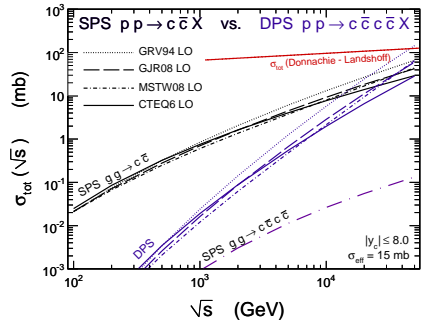}
   \caption{
\small Total LO cross section for single
$c \bar c$ pair and SPS and DPS $c \bar c c \bar c$ production
as a function of center-of-mass energy.  
}
 \label{fig:single_vs_double_LO}
\end{figure}

In Fig.~\ref{fig:single_vs_double_LO} we
compare cross sections for the single $c \bar c$ pair production as well
as for single-parton and double-parton scattering $c \bar c c \bar c$
production as a function of proton-proton center-of-mass energy. 
At low energies the conventional single $c \bar c$ pair production
cross section is much larger. 
The cross section for SPS production
of $c \bar c c \bar c$ system is more than two orders of magnitude smaller
than that for single $c \bar c$ production. For reference we show the
proton-proton total cross section as a function of energy.
At higher energies the DPS contribution
of $c \bar c c \bar c$ quickly approaches that for single $c \bar c$ 
production as well as the total cross section.

\begin{figure}[!h]
\includegraphics[width=5.0cm]{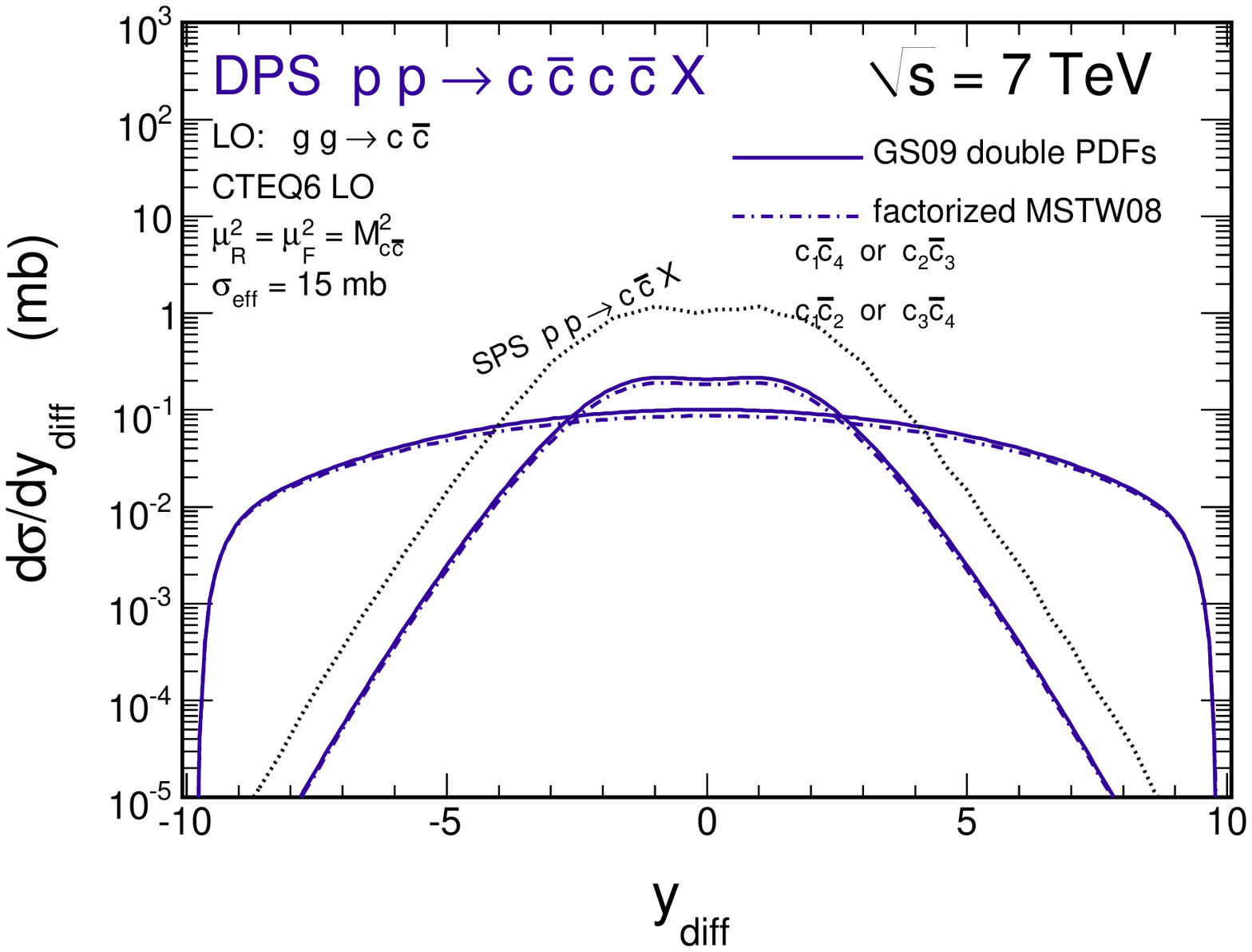}
\includegraphics[width=5.0cm]{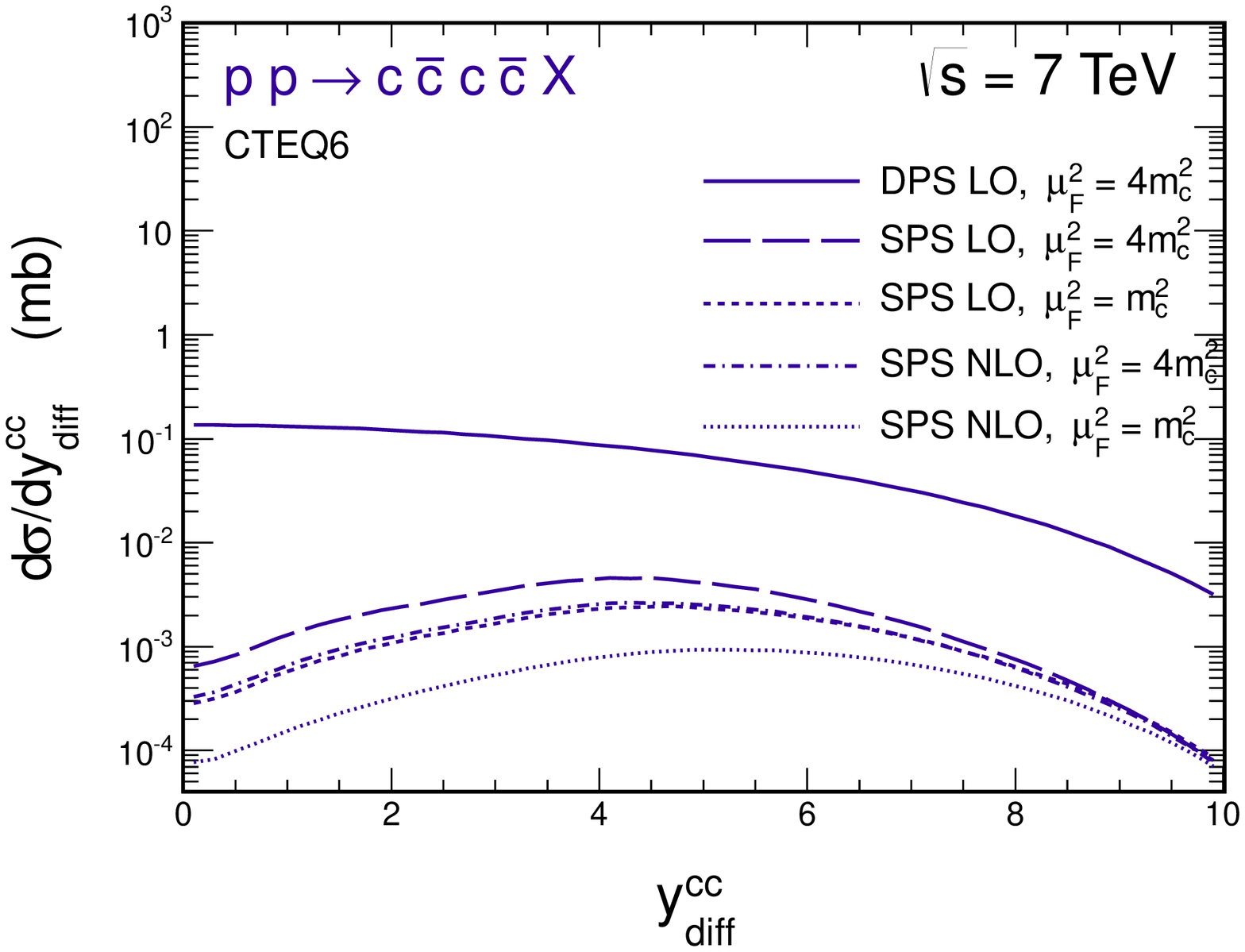}
   \caption{Comparison of SPS and DPS contributions. Left panel shows
results without and with QCD evolution of double parton distributions.
Right panel compares results for DPS and SPS production of 
$c \bar c c \bar c$.
}
 \label{fig:correlation_SPS_vs_DPS}
\end{figure}

In Fig.\ref{fig:correlation_SPS_vs_DPS} we show distributions
in rapidity difference between quark and antiquark from the same
scattering or between quarks from different scatterings (left panel). 
The distribution for $y_{cc}$ from different scatterings is much broader

than that for $c \bar c$ from the same scattering. In the right panel we
compare the SPS contribution with the DPS one. 


In Table 1 we show our first estimate of the cross sections for the
production of two $D$ mesons, both containing $c c$ quarks,
for different UGDF from the literature.
More details, including differential distributions, will be shown
in \cite{MS2012_DPS}. Our DPS estimate gives good order of magnitude
with respect to the LHCb data.

\begin{table}

\caption{Total cross sections for a production of pairs of mesons
within LHCb acceptance region.}
\begin{tabular}{|c|c|c|c|c|}
\hline
Mode & $\sigma_{tot}^{EXP}$ & KMR  & Jung setA$0+$ & KMS \\
\hline
  $D^{0}D^{0}$          & $690\pm40\pm70$   & 256   & 101  & 100 \\ 
  $D^{0}D^{+}$          & $520\pm80\pm70$   & 204   & 81   & 80  \\ 
  $D^{0}D^{+}_{S}$      & $270\pm50\pm40$   & 72    & 29   & 28  \\  
  $D^{+}D^{+}$          & $80\pm10\pm10$    & 41    & 16   & 16  \\
  $D^{+}D^{+}_{S}$      & $70\pm15\pm10$    & 29    & 12   & 11  \\
  $D^{+}_{S}D^{+}_{S}$  & $-$               & 10    & 4    & 4   \\
\hline
\end{tabular}


\end{table}

\section{Conclusions}

We have presented our selected new results for charmed meson production
at LHC. Results of our calculation have been compared with recent
ALICE and LHCb experimental data for transverse momentum distribution
of $D$ mesons. There seems to be a missing strength, especially
for the LHCb kinematics.

One of possible explanation is a presence of DPS contributions.
We have compared energy dependence of the DPS contribution to 
the $c \bar c c \bar c$ production with that for the $c \bar c$
production. The cross section for two pair production grows much faster
than that for single pair production. At high energies the two cross
sections become comparable. We have also discussed some correlation
observables that could be used to identify double-parton scattering
contribution. The rapidity difference between $cc$ (or $\bar c \bar c$)
is one of the good examples.

We have also estimated corresponding single-parton scattering
contributions in a high energy approach. The latter turned out
to be much smaller than the double-parton scattering contributions.

In Ref.\cite{LMS2011} we suggested that a good possibility to identify 
DPS effects would be to measure $D$ mesons of the same flavour. 
The LHCb collaboration has presented recently results of such first studies 
\cite{LHCb_DPS}. Our calculation predicts cross section of right order 
of magnitude.

In summary, we have found that the production of two $c \bar c$ pairs
is one of the best places to study and identify double-parton scattering
effects.

This work was supported in part by the MNiSW grant
No. PRO-2011/01/N/ST2/04116.



\bibliographystyle{aipproc}   

\bibliography{sample}

\IfFileExists{\jobname.bbl}{}
 {\typeout{}
  \typeout{******************************************}
  \typeout{** Please run "bibtex \jobname" to optain}
  \typeout{** the bibliography and then re-run LaTeX}
  \typeout{** twice to fix the references!}
  \typeout{******************************************}
  \typeout{}
 }


\end{document}